\documentclass[a4,12pt]{article}
\usepackage{amsmath}
\usepackage{amssymb}
\usepackage{graphicx}
\usepackage{color}
\usepackage{epsf}
\usepackage{colordvi}
\usepackage{epic}
\usepackage{bm}
\usepackage{fancybox} 

\begin{document}
\begin{titlepage}
\begin{flushright}
  KUNS-1960
\end{flushright}

\begin{center}
\vspace*{10mm}

{\LARGE \bf Flavor structure and coupling selection rule from 
intersecting D-branes }
\vspace{12mm}

{\large
Tetsutaro~Higaki\footnote{E-mail address: tetsu@gauge.scphys.kyoto-u.ac.jp},
Noriaki~Kitazawa\footnote{E-mail address: kitazawa@phys.metro-u.ac.jp},
Tatsuo~Kobayashi\footnote{E-mail address:
  kobayash@gauge.scphys.kyoto-u.ac.jp}
~and~~Kei-jiro~Takahashi\footnote{E-mail address:
keijiro@gauge.scphys.kyoto-u.ac.jp}
}
\vspace{6mm}

{\it $^{1,3,4}$Department of Physics, Kyoto University,
Kyoto 606-8502, Japan}\\[1mm]

{\it $^2$Department of Physics, Tokyo Metropolitan University,
Hachioji, Tokyo 192-0397, Japan}

\vspace*{15mm}

\begin{abstract}
We study flavor structure and the coupling selection rule 
in intersecting D-brane configurations.
We formulate the selection rule for Yukawa couplings and its 
extensions to generic n-point couplings. 
We investigate the possible flavor structure, which can 
appear from intersecting D-brane configuration, 
and it is found  that their couplings are determined by 
discrete abelian symmetries.
Our studies on the flavor structure and the 
coupling selection rule show that 
the minimal matter content of the supersymmetric 
standard model would have difficulty to 
derive realistic Yukawa matrices from stringy 3-point 
couplings at the tree-level.
However, extended models have a richer structure, 
leading to non-trivial mass matrices.

\end{abstract}

\end{center}
\end{titlepage}

\section{Introduction}

Understanding the origin of fermion masses and mixing 
angles is one of most important issues in particle physics.
Their experimental values show the hierarchical structure.
Within the framework of the standard model and 
its extensions, fermion masses are obtained through 
Yukawa couplings between fermions and Higgs fields.
Yukawa couplings seem  naturally of $O(1)$ in a sense. From 
this viewpoint, how to derive suppressed Yukawa couplings 
is a key-point for understanding hierarchical structure 
among fermion masses and mixing angles.

Superstring theory is a promising candidate for 
unified theory including gravity.
Thus, it is important to study which type of flavor structure 
can be derived in superstring theory.
Superstring  theory predicts the existence of 6D  
compact space in addition to our 4D spacetime.
Indeed, the 6D compact space is the origin of the flavor 
structure, 
that is, the flavor structures, which is derived from 
string models, are determined by geometrical aspects 
of the 6D compact space.

Several types of string  models have been proposed so far.
Recently, intersecting D-brane models have been studied extensively.
(See for essential idea on model building 
Refs.\cite{BGKL,AFIRU,Blumenhagen:2000ea,CSU}.)\footnote{See also
for  review ,
e.g. Refs. \cite{Kokorelis:2004tb,Blumenhagen:2005mu} 
and references therein.}
In this class of models, matter fields as well as Higgs fields 
correspond to open string modes at intersecting points 
between different D-branes, and those are localized modes 
\cite{Berkooz:1996km}.
Thus, Yukawa couplings among fermions and Higgs fields 
depend on distance of their intersecting points.
Suppressed Yukawa couplings can be realized when 
intersecting points corresponding to matter fermions and 
Higgs fields are far each other.
That is one of phenomenologically interesting aspects 
in intersecting D-brane models.
Such behavior is qualitatively the same as 
one in heterotic orbifold models. 
In general, an orbifold has fixed points on 
the 6D compact space, and 
heterotic orbifold models have twisted string modes, 
which are localized at orbifold fixed points.
Yukawa couplings can be suppressed when matter fermions and 
Higgs fields correspond to different fixed points.
Indeed, the same technique of conformal field theory (CFT) 
is used to calculate magnitudes of Yukawa couplings 
in both heterotic orbifold models 
\cite{Dixon:1986qv,Hamidi:1986vh,Burwick:1990tu,Kobayashi:2003vi} 
and intersecting D-brane models \cite{Cremades:2003qj,
Cvetic:2003ch,Abel:2003vv}.
(See also \cite{Cremades:2004wa}.)

Understanding the origin of suppressed Yukawa couplings 
is the first important step to derive realistic Yukawa matrices, 
but that is not sufficient to realize experimental values.
Derivation of realistic Yukawa matrices in string models 
is quite non-trivial.
Another important point is to study stringy selection rules 
for allowed Yukawa couplings.
Then we would see a pattern of Yukawa matrices in a model.
In heterotic orbifold  models, allowed Yukawa couplings 
are determined by the 6D space group selection rule 
\cite{Dixon:1986qv,Kobayashi:1991rp,Casas:1991ac} 
as well as $H$-momentum conservation and gauge invariance.
The space group selection rule is the 
selection rule for allowed Yukawa couplings due to 
the 6D orbifold geometry.
That is quite non-trivial, that is, 
in some types of Yukawa couplings on orbifolds 
only diagonal couplings are allowed, but 
for certain types of Yukawa couplings on 
non-prime orbifolds off-diagonal couplings are also 
allowed.
Here the diagonal couplings mean the case that 
when we choose two states, the other state to 
be allowed to couple is determined uniquely.
The off-diagonal couplings correspond to the case 
that the third states are not uniquely determined.
Obviously, off-diagonal Yukawa couplings are important  
to realize non-vanishing mixing angles in the case 
with the minimal and small numbers of Higgs fields.
Actually, the number of possible 6D orbifolds is finite 
as determined by 6D crystallographic space groups, 
and the number of fixed points on an orbifold is 
also finite.
The explicit results for allowed Yukawa couplings 
due to the space group selection rule are shown in 
\cite{Kobayashi:1991rp,Casas:1991ac}.
Then, in principle systematic study is possible to 
classify which patterns of Yukawa matrices can be obtained 
within the framework of heterotic orbifold models.
Indeed such analysis has been done for $Z_6$ orbifold models in 
\cite{Ko:2004ic} showing interesting results 
for the quark and lepton sectors.

On the other hand, patterns of Yukawa matrices have been 
calculated model by model explicitly 
within the framework of intersecting D-brane models so far.
Although explicit models with phenomenologically 
interesting aspects other than Yukawa matrices 
have been constructed so far, 
it is still a challenging issue to lead non-trivial Yukawa matrices.
Thus, our purpose is to study systematically flavor structure 
which can be derived from intersecting D-brane models, and 
classify possible patterns of Yukawa matrices.
For such purpose, we formulate selection rule for 
allowed Yukawa couplings in generic case.
In particular, we are interested in formulating 
the selection rule, which is useful to investigate 
whether off-diagonal couplings are allowed or not.
Indeed, the selection rule for allowed three-point 
couplings has been discussed in \cite{Cremades:2003qj}.
We will formulate the coupling selection rule in a different 
approach.
That is  similar to 
the space group selection rule in 
heterotic orbifold models, showing the condition for 
off-diagonal couplings.
Then we study which types of flavor structure can 
appear from intersecting D-brane configurations.
Here we concentrate intersecting D6-brane systems (as well as 
D4-brane systems) in type IIA string theory.
Since their T-duals correspond to magnetized D-brane systems in type IIB 
string theory, the following discussions would be applicable 
to magnetized D-brane systems in type IIB string theory.

This paper is organized as follows.
In section 2, we review briefly the space group selection 
rule in heterotic orbifold models.
In section 3, we study the selection rule in intersecting 
D-brane models.
In section 4, we consider which types of flavor structure 
can appear from intersecting D-brane configurations.
In subsection 4.1 we discuss the flavor structure, that 
the numbers of left-handed and right-handed quarks are 
the same on a $T^2$.
In this case, we show the minimum number of Higgs fields is 
the same as  the flavor number $N_f$, and 
the Higgs number is generically equal to 
$k N_f$ with $k \in {\bf N}$.
The selection rule is controlled by a discrete abelian symmetry.
In subsection 4.2, we discuss generic flavor structure, that is, 
the asymmetric flavor structure.
In that case, the coupling selection rule is also determined by 
a discrete abelian symmetry.
For the minimal number of Higgs field,   
we also show Yukawa matrices are non-trivial, but 
lead to a certain number of 
massless fermions, and diagonal entries are  larger than 
off-diagonal entries. 
In section 4.3, we comment on how to reduce the flavor and 
Higgs numbers, and that would be important to obtain 
non-trivial fermion mass matrices.
Section 5 is devoted to conclusion and discussion.
In Appendix A, we give a simple recipe: how to obtain 
shift vectors, which play an important role on 
description of independent intersecting points and 
formulation of the coupling selection rule.
In Appendix B, we discuss the possible number 
of Higgs fields when the flavor number of both left- and 
right-handed quarks on a $T^2$ is equal to two.

\section{Yukawa selection rule for 
heterotic orbifold models}

Here we give a brief review on the space group 
selection rule for allowed Yukawa couplings in 
heterotic orbifold models.
That will be useful to study the selection rule 
for allowed Yukawa couplings in intersecting D-brane 
models.

\subsection{Orbifold and fixed points}

An orbifold is obtained by dividing a torus $T^d$ by 
a twist $\theta$, while $T^d$ is a division of  
$R^d$ by a lattice $\Lambda$, i.e. $T^d = R^d/\Lambda$, 
where the twist $\theta$ must be automorphism of the 
lattice $\Lambda$.
On such orbifold, there are closed strings, which satisfy 
the following twisted  boundary condition,
\begin{equation}
X^i(\sigma = \pi ) = \left( \theta^k X(\sigma =0)\right )^i + v^i ,
\label{twist-bc}
\end{equation}
where $v^i$ is a shift vector on the torus lattice $\Lambda$.
These are refereed as the $\theta^k$ twisted string.
Their zero-modes $f^i$ also satisfy the same boundary condition, 
\begin{equation}
f^i =  (\theta^k f)^i + v^i ,
\end{equation}
that is the fixed point on the orbifold.
The fixed point $f^i$ is denoted by the corresponding 
space group element $(\theta^k,v^i)$.
The product of two space group elements, 
$(\theta^{k_{(1)}},v^i_{(1)})$ and $(\theta^{k_{(2)}},v^i_{(2)})$, 
is obtained as 
\begin{equation}
(\theta^{k_{(1)}},v^i_{(1)}) (\theta^{k_{(2)}},v^i_{(2)}) = 
(\theta^{k_{(1)}} \theta^{k_{(2)}},v^i_{(1)} + \theta^{k_{(1)}} v^i_{(2)}) .
\end{equation}
Indeed, this product of space group elements 
correspond to the combination of two twisted strings 
with the twisted boundary conditions, 
$(\theta^{k_{(1)}},v^i_{(1)})$ and $(\theta^{k_{(2)}},v^i_{(2)})$.

As said above, the fixed point and twisted string are denoted 
by the corresponding space group element $(\theta^k,v^i)$.
Furthermore, it is remarkable that the fixed point $f^i$ is 
equivalent to $f^i + \Lambda$ on the torus.
That implies that the space group $(\theta^k,v^i)$ 
is equivalent to $(\theta^k,v^i+ (1-\theta^k)\Lambda)$, 
that is, these belong to the same conjugacy class.
Thus, independent fixed points are defined up to 
the conjugacy classes.

Here we give two examples.
One is the 2D $Z_3$ orbifold, and the other is 
the 2D $Z_6$ orbifold.
The 2D $Z_6$ orbifold is obtained as a division of $R^2$ 
by the $SU(3)$ root lattice $\Lambda_{SU(3)}$ and its 
$Z_3$ automorphism $\theta$.
Here we denote the two simple roots of $\Lambda_{SU(3)}$ 
by $e_1$ and $e_2$, and these are transformed under $\theta$ 
as 
\begin{equation}
\theta e_1 = e_2, \qquad \theta e_2 = - e_1 - e_2.
\end{equation}
The lattice $(1-\theta)\Lambda$ is spanned e.g. by 
$3e_1$ and $e_1-e_2$.
The three independent fixed points are denoted by 
\begin{equation}
(\theta,ne_1), \qquad (n=0,1,2).
\end{equation}

Similarly, fixed points on the 2D $Z_6$ orbifold are 
obtained.
The 2D $Z_6$ orbifold is obtained as a division of $R^2$ 
by the $SU(3)$ lattice $\Lambda_{SU(3)}$ and the  
$Z_6$ twist, which transforms the $SU(3)$ simple roots as 
\begin{equation}
\theta e_1 = e_1 + e_2, \qquad \theta (e_1 +e_2) = e_2, 
\qquad \theta e_2 = -e_1 .
\end{equation}
Thus, we obtain $(1 - \theta) \Lambda = \Lambda$, 
that is, we have only one independent fixed point under $\theta$, 
i.e. 
\begin{equation}
(\theta,0) .
\end{equation}
The $\theta^2$ twist of $Z_6$ is nothing but the $Z_3$ twist.
Hence, the $\theta^2$ twist has three independent fixed points,
\begin{equation}
(\theta^2,ne_1), \qquad (n=0,1,2).
\end{equation}
Furthermore, the $\theta^3$ twist of $Z_6$ is the 
$Z_2$ twist, and the lattice $(1 - \theta^3)\Lambda$ is 
spanned by $2e_1$ and $2e_2$.
Thus, the four independent fixed points are denoted as 
\begin{equation}
(\theta^3,n_1e_1+n_2e_2), \qquad (n_i=0,1).
\end{equation}

Similarly, we can obtain fixed points on other orbifolds, 
and the twisted strings also correspond to those 
fixed points.\footnote{In non-prime order orbifold models, 
we have to take linear combinations of states corresponding 
directly to fixed points, in order to obtain 
$\theta$-eigenstates \cite{Kobayashi:1990mc}.
However, that is irrelevant to intersecting D-brane models.}
In the next subsection, we show the space group selection rule 
for these twisted strings.

\subsection{Space group selection rule}

In this subsection, we give a brief review on 
the space group selection rule for allowed Yukawa couplings.
Here we consider the condition that three twisted strings 
with the boundary conditions $(\theta^{k_{(a)}},v^i_{(a)})$ 
 for $a=1,2,3$  are allowed to couple.
The coupling condition due to the space group invariance 
is denoted as follows,
\begin{equation}
\prod_a (\theta^{k_{(a)}},v^i_{(a)}) = (1,0) .
\end{equation}
Simply, that implies the coupling is allowed  
when the three twisted strings combine into 
an untwisted  closed string, which can shrink.
However, here we recall that each space group element 
$(\theta^{k_{(a)}},v^i_{(a)})$ is equivalent to 
$(\theta^{k_{(a)}},v^i_{(a)}+(1 -\theta^{k_{(a)}})\Lambda ))$.
This equivalence in the conjugacy class has 
an important meaning, as shown below in explicit models.
The space group selection rule includes the point 
group selection rule, and the latter requires  
the product $\prod_a \theta^{k_{(a)}}$ to be identity.

Here we examine the space group selection rule for 
the two explicit models, the 2D $Z_3$ orbifold model and 
the 2D $Z_6$ orbifold model.
First, let us consider the 2D $Z_3$ orbifold models.
As shown in the previous subsection, there are three 
independent fixed points on the 2D $Z_3$ orbifold, 
$(\theta, ne_1$) with $n=0,1,2$.
Let us consider the coupling of three twisted strings 
corresponding to fixed points, $(\theta, n_1 e_1)$, 
$(\theta, n_2 e_1)$ and $(\theta, n_3 e_1)$.
The space group selection rule requires
\begin{equation}
(\theta, n_1 e_1)(\theta, n_2 e_1)(\theta, n_3 e_1) = (1,0) ,
\end{equation}
up to the conjugacy classes, and leads the following 
condition for allowed Yukawa couplings,
\begin{equation}
n_1 + n_2 + n_3 = 0 \quad \pmod{3}.
\end{equation}
That implies that the couplings are allowed only two cases, 
1) the case that  all of three fixed points are the same, and 
2) the case that all of three fixed points are different each other.
Namely, these are diagonal  couplings.
Indeed, this coupling selection rule can be understood as 
the $Z_3$ symmetry.

Similarly, we examine the space group selection rule on 
the 2D $Z_6$ orbifold.
As shown in the previous subsection, there are three twisted sectors, 
$\theta$-twisted, $\theta^2$-twisted  and $\theta^3$-twisted sectors.
For example, let us consider the couplings among 
$\theta$-twisted, $\theta^2$-twisted  and $\theta^3$-twisted sectors.
As shown in the previous subsection, the $\theta$-twisted,  
$\theta^2$-twisted  and $\theta^3$-twisted sectors
have one, three and four independent fixed points, which are 
denoted by $(\theta,0)$, $(\theta^2,ne_1)$ and 
$(\theta^3,m_1e_1+m_2e_2)$, respectively, 
where $n=0,1,2$ and $m_1,m_2=0,1$.
Now, we examine the condition for allowed Yukawa 
couplings due to the space group invariance, that is, 
\begin{equation}
(\theta,0)(\theta^2,ne_1)(\theta^3,m_1e_1+m_2e_2) = (1,0) ,
\end{equation}
where the space group elements are defined up to 
the conjugacy classes.
{\it The important point is that $(1-\theta)\Lambda = \Lambda$.
As a result, the couplings among all of the twisted states 
are allowed, and off-diagonal couplings are 
allowed.}
That makes it clear the situation that off-diagonal 
couplings are allowed, that is, 
when two or more independent fixed points under 
a twisted sector belong to the same conjugacy 
class in different twisted sector, off-diagonal 
couplings among corresponding twisted states are allowed.

The extension to the selection rule for generic $n$-point 
couplings is straightforward, that is, 
$\prod_{a = 1}^{n} (\theta^{k_{(a)}},v^i_{(a)}) = (1,0) $ 
for $n$ twisted strings with the boundary conditions, 
$(\theta^{k_{(a)}},v^i_{(a)})$ up to $(1-\theta^{k_{(a)}})\Lambda$.

\section{Coupling selection rule for intersecting D-brane models}

Here we study the coupling selection rule for 
intersecting D-brane models.
We concentrate intersecting D6-brane systems (as well as 
D4-brane systems) in type IIA string theory.
Since their T-duals may correspond to magnetized D-branes in type IIB 
string theory, the following discussions would be applicable 
to magnetized D-brane systems in type IIB string theory.
We consider $T^2\times T^2 \times T^2$ as the 6D compact space, 
and the D6-branes, which wrap one-cycle on each $T^2$.
Other backgrounds could be studied in a similar way, e.g. 
orientifold cases.
The $i$-th torus lattice $\Lambda^{(i)}$ is spanned by 
the basis $e_1^{(i)},e_2^{(i)}$.
Thus, the configuration of $D_a$-brane is described 
by its winding numbers for the $i$-th torus,
\begin{equation}
(n_a^{(i)},m_a^{(i)}) ,
\end{equation}
along $e_1^{(i)}$ and $e_2^{(i)}$.
Here we take $g.c.d.(n_a^{(i)},m_a^{(i)})=1$ or 
$(n_a^{(i)},m_a^{(i)}) = (1,0), (0,1)$. 
\footnote{
$g.c.d.(a,b)$ denotes the greatest common divisor of 
the integers $a$ and $b$.}
Namely, we consider the case that the vector 
${\bf w}^{(i)}_a =(n_a^{(i)},m_a^{(i)})$ is the shortest vector 
on the lattice $\Lambda^{(i)}$ along its direction.
A gauge multiplet appears on each set of D-branes.
The gauge group $U(N)$ is obtained from $N$ D-branes.

Now, let us consider two sets of D-branes, $D_a$-brane and 
$D_b$-brane, which intersect each other.
Their intersecting number on the $i$-th $T^2$ is 
given as 
\begin{equation}
I^{(i)}_{ab}=n_a^{(i)}m_b^{(i)} - n_b^{(i)}m_a^{(i)},
\end{equation}
and the total intersecting number on the 6D compact space is 
obtained as their product, i.e. 
$I_{ab} = I^{(1)}_{ab} I^{(2)}_{ab} I^{(3)}_{ab}$.
At these intersecting points, there appear open strings, 
one of whose ends is on the $D_a$-brane and the other 
is on the $D_b$-brane.
Such open strings can correspond to massless chiral matter 
fields, which have nontrivial representations under 
both gauge groups corresponding to 
$D_a$ and $D_b$ branes, that is, 
such massless modes have bi-fundamental representations 
$(N_a,\bar N_b)$ under $U(N_a) \times U(N_b)$ for positive 
intersecting numbers.
The different signs of $I_{ab}$ correspond to charge conjugated 
modes, i.e., $(\bar N_a,N_b)$.
Thus, chiral matter fields correspond to localized modes 
around intersecting points.
Higgs fields also correspond to such modes.
Here and hereafter we assume D-brane configurations with 
4D N=1 supersymmetry, because non-supersymmetric configurations are 
usually unstable.
Preserving supersymmetry requires certain conditions 
for intersecting angles, but that is irrelevant to our 
discussions.

Here, we study coupling selection rule among 
such open strings stretching different D-branes around 
intersecting points.
In heterotic orbifold models, the key-point for the coupling selection rule 
is description of boundary conditions of 
twisted strings (\ref{twist-bc}), that is, space group elements and 
their conjugacy classes.
Thus, let us study first how to describe boundary 
conditions of open strings around intersecting points.
Our setup is as follows.
We concentrate  a $T^2$ among $T^2 \times T^2 \times T^2$, 
e.g. the first torus, because extension to the case 
with $T^2 \times T^2 \times T^2$ is simple.
Then we consider two sets of D-branes, $D_a$ and $D_b$, 
whose winding numbers on the first $T^2$ are obtained as the 
following vectors on the torus lattice $\Lambda^{(1)}$,
\begin{eqnarray}
D_a &:& {\bf w}_a^{(1)} = (n_a^{(1)},m_a^{(1)}), \\
D_b &:& {\bf w}_b^{(1)} = (n_b^{(1)},m_b^{(1)}). 
\end{eqnarray}
Their intersecting number on the first $T^2$ is $I_{ab}^{(1)}$.
We assume the nontrivial case, i.e., $I_{ab}^{(1)} \neq 0$, and 
furthermore we take $I_{ab}^{(1)} > 0$, because the case with 
$I_{ab}^{(1)} < 0$ 
can be studied in the same way.
One of their intersecting points are taken as the origin of 
the torus lattice $\Lambda^{(1)}$.

Obviously, all of intersecting points sit along the 
$D_a$ brane, that is, those positions are written as 
$\frac{k}{I^{(1)}_{ab}} {\bf w}^{(1)}_a$ with 
$k=0,1,\cdots,I^{(1)}_{ab}-1$.
Note that the $D_b$ brane corresponding to $k=0$ passes 
the origin of the torus lattice, but the other $D_b$ branes 
corresponding to $k \neq 0$ do not.
On the other hand, all of intersecting points 
sit along $D_b$ brane, and those positions are 
written as 
$\frac{\ell}{I^{(1)}_{ab}} {\bf w}^{(1)}_b$ with 
$\ell=0,1,\cdots,I^{(1)}_{ab}-1$.
The former and the latter sets of independent 
intersecting points are equivalent each other on $T^2$.
The equivalence on $T^2$ implies the following relation,
\begin{equation}
\frac{k}{I^{(1)}_{ab}}  {\bf w}_a^{(1)} =  
\frac{\ell}{I^{(1)}_{ab}} {\bf w}_b^{(1)} + {\bf v}_{ab}^{(1)},
\end{equation}
for proper combinations of $k$ and $\ell$, 
where ${\bf v}_{ab}^{(1)}$ is a shift vector on the torus lattice.
Thus, the shift vectors ${\bf v}_{ab}^{(1)}$ can be used to 
describe independent intersecting points.
Note that $\left(\frac{k}{I^{(1)}_{ab}} + k'\right){\bf w}^{(1)}_a$
and $\left( \frac{\ell}{I^{(1)}_{ab}} + \ell' \right) {\bf w}^{(1)}_b$ 
are equivalent to $\frac{k}{I^{(1)}_{ab}} {\bf w}^{(1)}_a$ and 
$\frac{\ell}{I^{(1)}_{ab}} {\bf w}^{(1)}_b$, respectively, for  
$k',\ell' \in {\bf Z}$.
That implies that the shift vectors ${\bf v}_{ab}^{(1)}$ 
have the meaning to describe independent intersecting points 
up to the sublattice $\Lambda^{(1)}_{ab}$, which is 
spanned by ${\bf w}_a^{(1)}$ and ${\bf w}_b^{(1)}$.
Namely, the $I_{ab}^{(1)}$ independent shift vectors are   
coset representatives corresponding to $\Lambda^{(1)}/\Lambda^{(1)}_{ab}$.
The sublattice $\Lambda^{(1)}_{ab}$ is not as dense as 
$\Lambda^{(1)}$, \footnote{
Obviously we have $vol(\Lambda^{(1)}_{ab}) = 
I^{(1)}_{ab} vol(\Lambda^{(1)})$, where $vol(\Lambda)$ denotes the 
volume of the unit cell.}
and plays a role similar to $(1-\theta^{k})\Lambda$ in 
heterotic orbifold models.

Here we give a simple example.
We consider two sets of D-branes with winding numbers, 
${\bf w}^{(1)}_a=(1,0)$ and ${\bf w}^{(1)}_b=(1,3)$ as 
shown in Fig.~1.
Their intersecting number is $I^{(1)}_{ab}=3$.
Thus, the three independent intersecting points are 
written as $\frac{k}{3} {\bf w}^{(1)}_a$ with $k=0,1,2$.
Equivalent sets are obtained as 
$\frac{\ell}{3} {\bf w}^{(1)}_b$ with $\ell=0,1,2$.
On the other hand, the sublattice $\Lambda^{(1)}_{ab}$ can 
be spanned by (1,0) and (0,3), and that implies 
independent shift vectors ${\bf v}_{ab}^{(1)}$ up to 
$\Lambda^{(1)}_{ab}$ are obtained as ${\bf v}_{ab}^{(1)} = (0,-m)$ 
with $m = 0,1,2$.
Indeed, these intersecting points and shift vectors 
satisfy 
\begin{equation}
\frac{k}{3} {\bf w}^{(1)}_a = \frac{\ell}{3} {\bf w}^{(1)}_b + 
{\bf v}_{ab}^{(1)},
\end{equation}
up to $\Lambda^{(1)}_{ab}$, when $k=\ell=m$ as 
shown in Fig.~1.
Therefore, independent intersecting points 
and the coset representatives corresponding to 
$\Lambda^{(1)}/\Lambda^{(1)}_{ab}$ can 
be described by shift vectors ${\bf v}_{ab}^{(1)}$ up to 
$\Lambda_{ab}^{(1)}$.
See Appendix A, where a simple recipe how to obtain 
shift vectors ${\bf v}_{ab}^{(1)}$ for generic winding numbers is 
given.


\begin{figure}[htbp]
\epsfxsize=0.3\textwidth
\centerline{\epsfbox{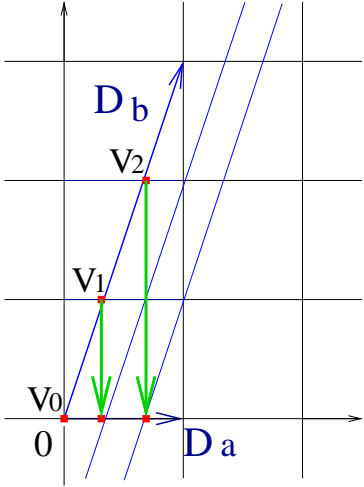}}
\caption{A simple example of D-brane configuration. 
The winding vectors ${\bf w}_a$ and ${\bf w}_b$ of 
$D_a:~(1,0)$ and $D_b:~(1,3)$ are shown.
The points on  ${\bf w}_a$ and ${\bf w}_b$ correspond to 
intersecting points $\frac{m}{3}{\bf w}_a$ and 
$\frac{m}{3}{\bf w}_b$ ($m=0,1,2$), respectively.
Shift vectors ${\bf v}_m=(0,-m)$ are also shown. 
}
\end{figure}


The above simple example also clarifies a concrete 
implication of shift vectors ${\bf v}_{ab}^{(1)}$ for open strings.
Let us consider the open string at the intersecting point 
$\frac{1}{3} {\bf w}^{(1)}_a$ between $D_a$ and $D_b$ branes, 
and move it such that one end sits on the origin 
of the torus lattice like Fig. 2.
The other end can not sit on the origin, but on the $D_b$ brane, 
which passes $(0,-1)$, that is nothing but the 
corresponding shift vector ${\bf v}_{ab}^{(1)}$.
The same is true for the other intersecting points.
Thus, we can describe open strings at different 
intersecting points by the following equations, 
which are satisfied by string end points, ${\bf x}_a$ and 
${\bf x}_b$,
\begin{equation}
{\bf x}_a - {\bf x}_b =  {\bf v}_{ab} ,
\label{open-bc}
\end{equation}
when we move one of end points to the same point, e.g. 
the origin.
It is also true for generic case, that is, 
independent intersecting points are represented by 
shift vectors ${\bf v}_{ab}$ up to $\Lambda_{ab}$, and 
end points of $D_a$-$D_b$ open string at the intersecting point  
satisfy eq.(\ref{open-bc}).


\begin{figure}[htbp]
\epsfxsize=0.3\textwidth
\centerline{\epsfbox{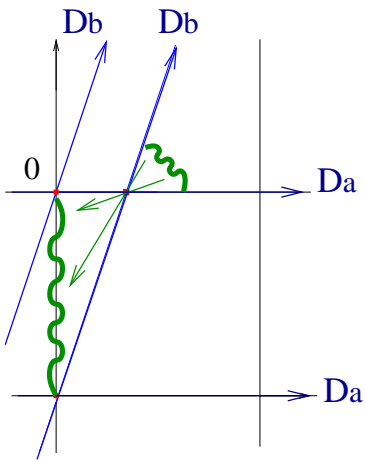}}
\caption{Open string at the intersecting point $\frac{1}{3}{\bf w}_a$.
We can move its end pont on the $D_a$-brane to the origin, and 
the other end point can be moved not to the origin, but the point 
$(0,-1)$.
}
\end{figure}


Now we are ready to study the coupling selection rule 
in intersecting D-brane models.
We consider three sets of D-branes, $D_a$, $D_b$ and $D_c$, 
with the following winding numbers,
\begin{eqnarray}
D_a &:& {\bf w}_a^{(1)} = (n_a^{(1)},m_a^{(1)}),\nonumber  \\
D_b &:& {\bf w}_b^{(1)} = (n_b^{(1)},m_b^{(1)}), \\
D_c &:& {\bf w}_c^{(1)} = (n_c^{(1)},m_c^{(1)}).  \nonumber
\end{eqnarray}
Here we do not consider the trivial case that 
one of $I_{ab}^{(1)}$, $I_{bc}^{(1)}$ and $I_{ca}^{(1)}$ vanishes.
Furthermore, we study the case that all of 
$I_{ab}^{(1)}$, $I_{bc}^{(1)}$ and $I_{ca}^{(1)}$ are positive.
It is quite straightforward to extend the following discussions 
to other cases, e.g. the case that all of 
$I_{ab}^{(1)}$, $I_{bc}^{(1)}$ and $I_{ca}^{(1)}$ are
negative.\footnote{We will show at the end of the section that 
the $H$-momentum conservation requires all of 
$I_{ab}^{(1)}$, $I_{bc}^{(1)}$ and $I_{ca}^{(1)}$ to have 
the same sign.}
There are three types of open strings, 
$D_a$-$D_b$, $D_b$-$D_c$ and $D_c$-$D_a$ open strings 
at intersecting points between $D_a$ and $D_b$ branes, 
$D_b$ and $D_c$ branes, and $D_c$ and $D_a$ branes, 
respectively.
Let us study the coupling selection rule among 
these three open strings.
Simply, these open strings can couple if 
the corresponding intersecting points and D-branes make 
a closed triangle.
In other words, the coupling is allowed if 
these three open strings combine into a closed string, 
which can shrink without winding on the torus.
Recall that the end points of the open string satisfy the 
condition (\ref{open-bc}).
Thus, the above condition for allowed couplings is written as 
\begin{equation}
{\bf v}_{ab}^{(1)} + {\bf v}_{bc}^{(1)} + {\bf v}_{ca}^{(1)} = 0 .
\label{selection-rule}
\end{equation}
Here note that shift vectors, ${\bf v}_{ab}^{(1)}$, ${\bf v}_{bc}^{(1)}$ and 
${\bf v}_{ca}^{(1)}$, are defined up to the sublattices, 
$\Lambda_{ab}^{(1)}$, $\Lambda_{bc}^{(1)}$ and $\Lambda_{ca}^{(1)}$, 
respectively.
This selection rule tells us whether off-diagonal couplings 
are allowed or not in intersecting D-brane models like 
heterotic orbifold models.
{\it For example, in the situation that differences between 
two or more independent shifts ${\bf v}_{ab}^{(1)}$ are on the sublattice 
$\Lambda_{bc}^{(1)}$, off-diagonal couplings are allowed.}
Thus, the difference among the sublattices, 
$\Lambda_{ab}^{(1)}$, $\Lambda_{bc}^{(1)}$ and $\Lambda_{ca}^{(1)}$, 
is important to realize off-diagonal couplings.

As an illustrating example, let us consider 
D-brane configuration with the following winding numbers 
\begin{eqnarray}
D_a&:& \quad (1,0), \nonumber \\
D_b&:& \quad (-1,2), \label{ex-2} \\
D_c&:& \quad (1,-3). \nonumber
\end{eqnarray}
The intersecting number between $D_a$ and $D_b$ branes is 
obtained as $I_{ab}^{(1)}=2$, and independent intersecting points are 
described by 
\begin{equation}
{\bf v}_{ab}^{(1)} = (0,\ell), \qquad (\ell = 0,1),
\end{equation}
up to the sulattice $\Lambda_{ab}^{(1)}$, which is spanned by 
(1,0) and (0,2).
Similarly, the intersecting number between $D_c$ and $D_a$ branes is 
obtained as $I_{ca}^{(1)}=3$, and independent intersecting points are 
described by 
\begin{equation}
{\bf v}_{ca}^{(1)} = (0,k), \qquad (k = 0,1,2),
\end{equation}
up to the sublattice $\Lambda_{ca}^{(1)}$, which is spanned by 
(1,0) and (0,3).
On the other hand, the intersecting number between $D_b$ and 
$D_c$ is equal to  $I_{bc}^{(1)}=1$, and 
the corresponding shift vector ${\bf v}_{bc}^{(1)}$ is obviously obtained 
as ${\bf v}_{bc}^{(1)} = 0$ up to the sublattice $\Lambda_{bc}^{(1)}$, 
where $\Lambda_{bc}^{(1)}$ is nothing but the torus lattice, 
i.e. $\Lambda_{bc}^{(1)} = \Lambda^{(1)}$.
In this model, couplings among all the states are allowed.
More explicitly, the following Yukawa couplings,
\begin{equation}
Y_{\ell k} C_{(ab)}^\ell C_{(ca)}^k C_{(bc)},
\end{equation}
are allowed for any $\ell$ and $k$, 
where $C_{(ab)}^\ell$ denote chiral matter fields corresponding 
to the $D_a$-$D_b$ open strings and the other notation of fields, 
$ C_{(ca)}^k$ and $C_{(bc)}$, have a similar meaning.
Here we have assumed that all of intersecting numbers 
on the second and third tori are equal to one.
The strength of Yukawa couplings is calculated through 
CFT technique in 
\cite{Cremades:2003qj, Cvetic:2003ch, Abel:2003vv}, and 
its dominant factor is obtained as 
\begin{equation}
Y \sim e^{-S_{cl}} \sim e^{-(\Sigma_1 + \Sigma_2+\Sigma_3)},
\end{equation}
where $S_{cl}$ denotes the action of the classical string 
solution $X_{cl}$, which has the asymptotic behavior 
corresponding to local open string modes near intersecting points, 
and that is obtained as the triangle area $\Sigma_i$, which string 
sweeps to couple on the $i$-th $T^2$.
Here, we consider only the dominant contribution due to the 
minimal classical action, although other classical solutions 
also have sub-dominant contributions.
That is because we are interested in the hierarchical form of 
Yukawa matrices.
It is possible to extend the following discussions on 
possible forms of Yukawa matrices to full results including 
all of classical solutions \cite{Cremades:2003qj, Cvetic:2003ch, Abel:2003vv}.

For example, in the configuration that three sets of D-branes, 
$D_a$, $D_b$ and $D_c$, intersect at the same point, ``the origin'' of
the torus lattice, 
we evaluate    
\begin{equation}
Y_{\ell k} = \left(
\begin{array}{ccc}
1 & \varepsilon^4  & \varepsilon^4 \\ 
\varepsilon^9 & \varepsilon &  \varepsilon 
\end{array}
\right) .
\label{asym-1}
\end{equation}
Here, the parameter $\varepsilon$ denotes the 
suppression factor for the Yukawa coupling corresponding 
to the minimum triangle.
For explicit calculations of Yukawa couplings, 
it is simple to use a figure like Fig. 3, in particular 
in the case that the number of either Higgs field or matter field 
is equal to one.
Fig. 3 shows the above D-brane configuration, where 
the single $C_{(bc)}$ field is located at the origin.
Several vertical branes correspond to $D_a$ branes.
Intersecting points between $D_a$ and $D_b$ are labeled by 
integer $\ell$ (mod 2), while intersecting points between 
$D_a$ and $D_c$ are labeled by integer $k$ (mod 3). 
Indeed, off-diagonal couplings are allowed, 
because all of combinations $(\ell,k)$ can be connected 
by the $D_a$ brane in Fig.~3.
If $I^{(1)}_{ab}$ and $I_{ca}^{(1)}$ had 
$g.c.d.(I^{(1)}_{ab},I_{ca}^{(1)}) \neq 1$, all of 
combinations $(\ell,k)$ could not be connected by 
the $D_a$ brane in a figure like Fig.~3.


\begin{figure}[htbp]
\epsfxsize=0.6\textwidth
\centerline{\epsfbox{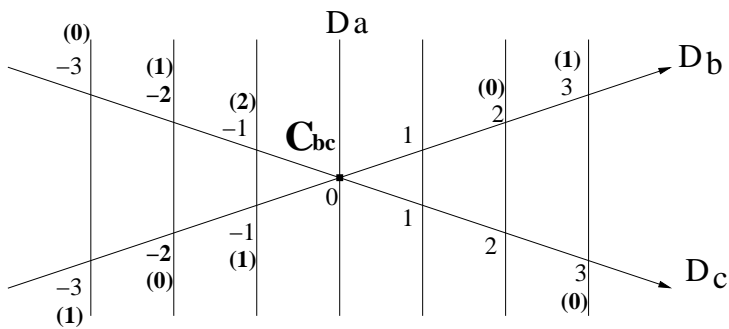}}
\caption{D-brane configuration corresponding to eqs.~(\ref{ex-2}), 
(\ref{asym-1}).
The single $C_{(bc)}$ field is located at the origin.
Intersecting points between $D_a$ and $D_b$ are labeled by 
integer $\ell$ (mod 2), while intersecting points between 
$D_a$ and $D_c$ are labeled by integer $k$ (mod 3). 
The parameter $\varepsilon$ in eq.~(\ref{asym-1}) denotes the 
suppression factor for the Yukawa coupling corresponding 
to the minimum triangle.}
\end{figure}


Even if the three D-branes do not intersect at the same point like 
Fig. 4, the coupling selection rule is the same, because the 
shift vectors do not change.
However, Yukawa matrix becomes 
\begin{equation}
Y_{\ell k} = \left(
\begin{array}{ccc}
\varepsilon^{d^2} & \varepsilon^{(2+d)^2}  & \varepsilon^{(2-d)^2} \\ 
\varepsilon^{(3-d)^2} & \varepsilon^{(1-d)^2} &  \varepsilon^{(1+d)^2} 
\end{array}
\right) ,
\label{asym-2}
\end{equation}
where $d$ is a continuous parameter ($-1 \leq d \leq 1$).


\begin{figure}[htbp]
\epsfxsize=0.6\textwidth
\centerline{\epsfbox{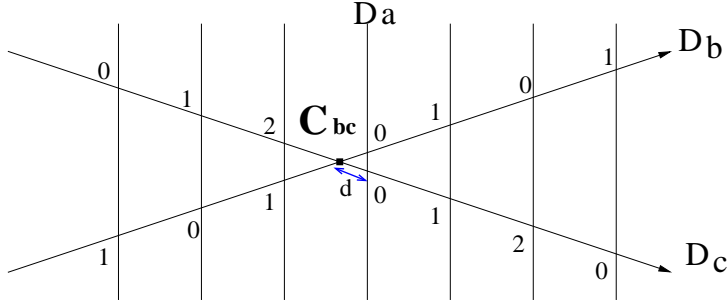}}
\caption{The same D-brane configuration as Fig.~3.
Three sets of  D-branes do not intersect at  the same point, and 
the distance between those intersecting points is parameterized by 
$d$.
}
\end{figure}


This example shows an important aspect of the coupling selection rule, 
that is,  all of couplings are allowed for any D-brane configuration with
\begin{equation}
g.c.d.(I_{ab}^{(1)},I_{bc}^{(1)})=g.c.d.(I_{ca}^{(1)},I_{bc}^{(1)})=
g.c.d.(I_{ab}^{(1)},I_{ca}^{(1)})=1.
\label{co-prime}
\end{equation}
This rule can be understood simply by drawing a figure like 
Fig. 3, by which we can see all of combinations of intersecting points 
are connected by the same type of $D$-branes.
That can also be explained as follows.
Suppose that 
\begin{equation}
g.c.d.(n_1^{(1)},n_2^{(1)})=g.c.d.(n_1^{(1)},n_3^{(1)})=
g.c.d.(n_2^{(1)},n_3^{(1)})=1.
\end{equation}
Then, the sublattice $\Lambda_{ab}^{(1)}$ is spanned by 
$(1,J_{ab}^{(1)})$ and $(0,I_{ab}^{(1)})$, and the sublattice 
$\Lambda_{ca}^{(1)}$ 
is spanned by $(1,J_{ca}^{(1)})$ and $(0,I_{ca}^{(1)})$, 
where $J_{ab}^{(1)}$ and $J_{ca}^{(1)}$ are certain integers, 
but irrelevant to our discussions.
(See Appendix A.1.)
Similarly, the sublattice $\Lambda_{bc}^{(1)}$ is spanned 
by $(1,J_{bc}^{(1)})$ and $(0,I_{bc}^{(1)})$.
It is obvious that the combination among the sublattices 
$\Lambda_{ab}^{(1)}, \Lambda_{ca}^{(1)}$ and $\Lambda_{bc}^{(1)}$ 
corresponds to the torus lattice $\Lambda^{(1)}$, i.e. 
$\Lambda^{(1)} = \Lambda_{ab}^{(1)} \cup \Lambda_{ca}^{(1)} \cup 
\Lambda_{bc}^{(1)}$,   
when eq.(\ref{co-prime}) is satisfied.
That implies that  all of couplings are allowed for any D-brane 
configuration with eq.(\ref{co-prime}).
In generic winding numbers, the sublattice $\Lambda_{ab}^{(1)}$ is 
spanned by $(k_{ab},J_{ab}^{(1)})$ and $(0,k'_{ab})$, and 
the sublattice $\Lambda_{ca}^{(1)}$ is spanned by 
$(k_{ca},J_{ca}^{(1)})$ and $(0,k'_{ca})$. (See Appendix A.2.)
Moreover, the lattice $\Lambda_{bc}^{(1)}$ is spanned by 
$(k_{bc},J_{bc}^{(1)})$ and $(0,k'_{bc})$.
Eq.(\ref{co-prime}) implies that 
\begin{eqnarray}
& & g.c.d.(k_{ab},k_{ca})=g.c.d.(k_{ab},k_{bc})=
g.c.d.(k_{ca},k_{bc})=1, \nonumber \\
& & g.c.d.(k'_{ab},k'_{ca})=g.c.d.(k'_{ab},k'_{bc})=
g.c.d.(k'_{ca},k'_{bc})=1 ,
\end{eqnarray}
where $I_{ab}^{(1)}=k_{ab}k'_{ab}$, $I_{ca}^{(1)}=k_{ca}k'_{ca}$ and 
$I_{bc}^{(1)}=k_{bc}k'_{bc}$.
That leads to the fact that the combination among the sublattices 
$\Lambda_{ab}^{(1)}, \Lambda_{ca}^{(1)}$ and $\Lambda_{bc}^{(1)}$ 
corresponds to the torus lattice $\Lambda^{(1)}$, i.e. 
$\Lambda^{(1)} = \Lambda_{ab}^{(1)} \cup \Lambda_{ca}^{(1)} \cup 
\Lambda_{bc}^{(1)}$.
That implies that  all of couplings are allowed for any D-brane 
configuration with eq.(\ref{co-prime}), 
and that has been shown already in \cite{Cremades:2003qj} 
through a different approach.
This rule will be extended into generic case when we 
find generic D-brane configurations in subsection 4.2.

The extension to the selection rule for generic $n$-point couplings 
is straightforward.
We consider $n$ sets of $D_i$ branes for $i=1,2,\cdots, n$.
Such setup may include open strings at intersecting points 
between $D_i$ and $D_{i+1}$ 
branes as well as open strings between $D_n$ and $D_1$ branes.
Their intersecting points are described by shift vectors 
${\bf v}_{i,i+1}$ as well as ${\bf v}_{n,1}$.
Then, the condition for allowed couplings is written as  
\begin{equation}
{\bf v}_{1,2} + {\bf v}_{2,3} + \cdots + {\bf v}_{n-1,n} + {\bf v}_{n,1} = 0.
\end{equation}
Here, recall that the shift vectors ${\bf v}_{i,i+1}$ are 
defined up to the lattice $\Lambda_{i,i+1}$, whose 
definition is the same as $\Lambda_{ab}$.

Finally we comment on the coupling selection rule due to 
the $H$-momentum conservation.
Within the bosonized formulation, the (twisted) 
RNS fermionic strings are written as
\begin{equation}
e^{iq_i H_i},
\end{equation}
where $H_i$ are 2D bosonized fields and $q_i$ are  
the so-called $H$-momenta.
If $I^{(i)}_{ab} > 0$, massless space-time spinors corresponding to 
R modes and massless space-time scalars corresponding to 
NS modes have the following $H$-momenta
\begin{eqnarray}
q_i &=& \theta_{ab}^{(i)} - \frac{1}{2} {\rm ~~for~~R~~}, \\
q_i &=& \theta_{ab}^{(i)} - 1 {\rm ~~for~~NS~~},
\label{H-momenta}
\end{eqnarray}
respectively,
where $\theta_{ab}^{(i)}\pi$ ($0 < \theta_{ab}^{(i)} < 1 $) 
denotes the intersecting angle 
on the $i$-th $T^2$ between $D_a$ and $D_b$ branes.
Now, let us consider the $H$-momentum conservation for 
Yukawa couplings among two fermions and a single scalar field, 
which are originated from $D_a$-$D_b$, $D_b$-$D_c$ and 
$D_c$-$D_a$ open strings.
The $H$-momentum conservation requires that 
\begin{equation}
(\theta_{ab}^{(i)} + \theta_{bc}^{(i)} + \theta_{ca}^{(i)}) \pi 
= 2 \pi.
\end{equation}
That implies that the sum of exterior angles 
of triangle must be equal to $2 \pi$.
Obviously, that is satisfied with a closed triangle.
However, if one of intersecting numbers, e.g. $I_{ab}^{(1)}$, is 
negative, the massless mode has the $H$-momentum opposite to
eq.(\ref{H-momenta}).
In this case, the $H$-momentum conservation must not 
be satisfied with a closed triangle configuration of 
$D$-branes.
Moreover, if all of intersecting numbers are negative, 
the $H$-momentum conservation is satisfied with a closed 
triangle configuration of D-branes.
Thus, the $H$-momentum conservation is satisfied only when 
all of intersecting numbers, $I_{ab}^{(i)}$,  $I_{bc}^{(i)}$ 
and $I_{ca}^{(i)}$, have the same sign on each $i$-th $T^2$.
That includes the condition that all of total intersecting 
numbers, $I_{ab}$,  $I_{bc}$ and $I_{ca}$, must have the same sign.
Since that means the corresponding Yukawa couplings are 
$(N_a,\bar N_b,1)(1,N_b,\bar N_c)(\bar N_a,1,N_c)$ type or its 
conjugate, the condition due to $H$-momentum conservation includes  
the condition that the Yukawa couplings must be gauge invariant.

\section{Flavor structure from intersecting D-brane 
configurations}

In this section, we study systematically flavor structures, 
which can appear from 
intersecting D-brane configurations.
The flavor structure on $T^2 \times T^2 \times T^2$ is a 
direct product of the flavor structure on each $T^2$.
Thus, we mainly investigate the flavor structures from 
intersecting D-brane configurations on each $T^2$.
In subsection 4.1, we study the symmetric flavor structure, 
that is, the intersecting numbers for left and right-handed 
quarks are the same on $T^2$.
In subsection 4.2, we study the asymmetric flavor structure, 
that is, the intersecting numbers for left and right-handed 
quarks are different each other on $T^2$, but their total 
numbers on $T^2 \times T^2 \times T^2$ are the same.

\subsection{Symmetric flavor structure}

Here we investigate the D-brane configurations leading to 
the symmetric flavor structure, 
that is, the intersecting numbers for left and right-handed 
quarks are the same on $T^2$.
Suppose that the flavor number is equal to $N_f$, e.g. on 
the first $T^2$ while their intersecting numbers on the 
other $T^2 \times T^2$ are equal to one.
Then, easily we can show the number of Higgs fields, which can 
have allowed 3-point couplings with these quarks, is equal to 
$k N_f$, where $k \in {\bf Z}$ and $N_f \neq 2$.
Let us consider three sets of D-branes, $D_C$, $D_L$ and $D_R$ branes 
with the following winding numbers,
\begin{eqnarray}
D_C &:& {\bf w}_C = (n_C,m_C),\nonumber  \\
D_L &:& {\bf w}_L = (n_L,m_L), \\
D_R &:& {\bf w}_R = (n_R,m_R).  \nonumber
\end{eqnarray}
In this subsection, we omit the index $(1)$ denoting 
the first $T^2$, because we discuss the D-brane configurations 
only on the first $T^2$.
Open strings between $D_C$ and $D_L$ $(D_R)$ correspond to 
left-handed (right-handed) quarks $Q_L$ ($Q_R$), and 
open strings between $D_L$ and $D_R$ correspond to 
modes, which can have allowed 3-point couplings with 
left and right-handed quarks, that is, 
Higgs fields $H$.
Then, we consider the symmetric flavor structure,
\begin{equation}
|I_{CL}|=|I_{RC}|=N_f.
\end{equation}
Through a simple algebraic calculation, that implies that  
\begin{equation}
{\bf w}_L \pm {\bf w}_R = k {\bf w}_C ,
\end{equation}
where $k$ is a real number.
The sign in l.h.s. depends on the signs of 
$I_{CL}$ and $I_{RC}$.\footnote{The $H$-momentum conservation 
and gauge invariance require $I_{CL}=I_{RC}$, and 
in this case, the sign in l.h.s. must be $+$.}
Recall here that  
${\bf w}_C$ is the shortest vector on $\Lambda$ 
along its direction.
That implies that $k$ must be integer.
Thus, we obtain the number of Higgs fields 
\begin{equation}
|I_{LR}|=k N_f.
\end{equation}
Namely, the minimum number of Higgs fields is equal to 
$N_f$.\footnote{In the case with $N_f=2$, the minimum number of 
Higgs is not equal to 2, but 4. See Appendix B for such case.}
In this case, we find  
\begin{equation}
\Lambda_{CL}=\Lambda_{RC}=\Lambda_{LR},
\end{equation}
and the varieties of shift vectors, ${\bf v}_{CL}$, ${\bf v}_{RC}$ and 
${\bf v}_{LR}$ are the same.
Thus, this type of D-brane configurations 
lead to only diagonal couplings for one of $N_f$ Higgs fields.
Indeed, the selection rule is determined by the 
discrete abelian symmetry.

As an illustrating example, we consider the case of $N_f=3$, 
e.g. with the following winding numbers,
\begin{eqnarray}
D_C &:& {\bf w}_C=(n_C,m_C) = (1,0),  \nonumber \\
D_L &:& {\bf w}_L=(n_L,m_L) = (1,3),  \\
D_R &:& {\bf w}_R=(n_R,m_R) = (-2,-3). \nonumber  
\end{eqnarray}
Indeed, this configuration leads to 
$I_{CL}=I_{RC}=I_{LR}=3$.
All of the sublattices, 
$\Lambda_{CL}$, $\Lambda_{RC}$ and $\Lambda_{LR}$, 
are the same, and spanned by (1,0) and (0,3).
The intersecting points of these D-branes are 
described by the shift vectors,
\begin{equation}
{\bf v}_{CL}=(0,k_{CL}), \qquad {\bf v}_{RC}=(0,k_{RC}), \qquad 
{\bf v}_{LR}=(0,k_{LR}), 
\end{equation}
where $k_{CL}, k_{RC}, k_{LR}=0,1,2$ (mod 3).
Thus, the coupling selection rule (\ref{selection-rule}) 
leads to 
\begin{equation}
k_{CL} + k_{RC}+ k_{LR}=0, \qquad \pmod{3}.
\end{equation}
This selection rule is the same as one in the 2D $Z_3$ 
orbifold, and is described by the $Z_3$ symmetry, under which 
the fields $\Phi_k$ with the $Z_3$ charge $k$ transform as 
\begin{equation}
\Phi_k \rightarrow e^{2\pi i k/3} \Phi_k,
\end{equation}
where $k=0,1,2$.
Thus, only diagonal couplings are allowed for one of three 
Higgs fields.
Explicitly, in the case that the three D-branes intersect 
at the same point, we obtain the Yukawa matrix,
\begin{equation}
Y = \left(
\begin{array}{ccc}
H_0 & \varepsilon H_2 & \varepsilon H_1 \\
\varepsilon H_2 & H_1 & \varepsilon H_0 \\
\varepsilon H_1 & \varepsilon H_0 & H_2
\end{array}
\right),
\end{equation}
where $\varepsilon$ is the suppression factor.
This form of Yukawa coupling is generic form when 
the three sets of D-branes intersect at the same point, 
although we have shown the explicit winding numbers.
When $\varepsilon$ is sufficiently suppressed, 
its three eigenvalues are obtained by vacuum expectation values 
(VEVs) of the Higgs fields,  
$v_i = \langle H_i \rangle $, for $i=0,1,2$, 
and its diagonalizing matrix is obtained as\footnote{
See e.g. \cite{Abel:2002ih} for similar analysis on $Z_3$ orbifold 
models.}
\begin{equation}
V = \left(
\begin{array}{ccc}
1 & -\frac{v_2}{v_1} \varepsilon & -\frac{v_1}{v_2} \varepsilon  \\
\frac{v_2}{v_1} \varepsilon & 1 & -\frac{v_0}{v_2} \varepsilon  \\
\frac{v_1}{v_2} \varepsilon & \frac{v_0}{v_2} \varepsilon  & 1
\end{array}
\right) .
\end{equation}

We apply the above Yukawa matrix to the up and down sectors 
of quarks with Higgs fields $H^{(u,d)}_i$ and suppression 
factors $\varepsilon_{u,d}$.
The quark mass ratios are  obtained as 
\begin{equation}
m_u : m_c : m_t = v^u_0 : v^u_1 : v^u_2, \qquad 
m_d : m_s : m_b = v^d_0 : v^d_1 : v^d_2.
\end{equation}
Moreover, the mixing angles are predicted as 
\begin{eqnarray}
V_{us} &=& \frac{m_b}{m_s}\varepsilon_d -
\frac{m_t}{m_c}\varepsilon_u, 
\nonumber \\
V_{ub} &=& \frac{m_s}{m_b}\varepsilon_d -  \frac{m_c}{m_t}\varepsilon_u, \\
V_{cb} &=& \frac{m_d}{m_b}\varepsilon_d -  \frac{m_u}{m_t}\varepsilon_u ,
\nonumber
\end{eqnarray}
by the two parameters $\varepsilon_{u,d}$.
This prediction does not fit the experimental values.

When the three sets of D-branes do not intersect at 
the same point, the Yukawa matrix becomes 
\begin{equation}
Y = \left(
\begin{array}{ccc}
\varepsilon^{d^2} H_0 & \varepsilon^{(1-d)^2} H_2 & \varepsilon^{(1+d)^2} H_1 \\
\varepsilon^{(1+d)^2} H_2 & \varepsilon^{d^2}H_1 & \varepsilon^{(1-d)^2} H_0 \\
\varepsilon^{(1-d)^2} H_1 & \varepsilon^{(1+d)^2} H_0 & \varepsilon^{d^2}H_2
\end{array}
\right),
\end{equation}
where $d$ is a continuous parameter $(-1 \leq d \leq 1)$ 
to denote the nearest distance among
three types of intersecting points.
This is generic form of Yukawa matrix for the 
symmetric flavor structure with $N_f=3$ flavor 
and the three Higgs fields.
Application of this form to the up and down sectors of 
quarks does not seem to lead to fully realistic 
results \cite{Chamoun:2003pf}.

Similarly we can discuss the D-brane configurations 
with more than three Higgs fields, i.e. the case with 
the $N_f k$ Higgs fields for $k>1$.
In this case, the sublattices $\Lambda_{CL}$ and $\Lambda_{RC}$ 
are still the same and these are spanned by ${\bf w}_C$ and 
${\bf w}_L$.
On the other hand, the sublattice $\Lambda_{LR}$ is 
spanned by $k{\bf w}_C$ and ${\bf w}_L$, and is less 
dense than $\Lambda_{CL}$ and $\Lambda_{RC}$.
That implies the number of coset representatives corresponding to 
$\Lambda/\Lambda_{LR}$ 
is $k$ times as large as one of $\Lambda/\Lambda_{CL}$ and 
$\Lambda/\Lambda_{RC}$, 
that is, the independent set of intersecting 
points for the Higgs fields is described by the joint of   
sets of shift vectors $\{ {\bf v}_{CL} \} \cup \{ m{\bf w}_C\}$ for 
$m=0,1,\cdots, (k-1)$, 
while the independent set of intersecting 
points for left and right-handed quarks is 
described by the set of shift vectors $\{ {\bf v}_{CL} \}$.
However, the part $\{ m{\bf w}_C \}$ is irrelevant to the 
coupling selection rule, 
that is, the selection rule is determined by the 
same $Z_{N_f}$ symmetry, and the part   $\{ m{\bf w}_C \}$ has 
the trivial charge under the $Z_{N_f}$ symmetry.
As a result, the same couplings are allowed except replacing 
\begin{equation}
H({\bf v}_{ab}) \rightarrow H({\bf v}_{ab}) +H({\bf v}_{ab}+{\bf w}_C) 
+ \cdots  +  H({\bf v}_{ab}+(k-1){\bf w}_C ),
\end{equation}
where 
$H({\bf v}_{ab})$ denotes the Higgs field corresponding to 
the shift vector ${\bf v}_{ab}$ and 
we have omitted  coefficients.
As a result, only diagonal couplings are allowed for 
one of Higgs fields.

Here we give an explicit model with $N_f$ flavors and $2N_f$ 
Higgs fields.
In this case we have the following Yukawa matrix,
\begin{equation}
Y=\left(
\begin{array}{ccc}
H_1 + \varepsilon^9 H_5 & \varepsilon H_4 + \varepsilon^4 H_3 & 
\varepsilon H_6 + \varepsilon^4 H_2 \\
\varepsilon H_4 + \varepsilon^4 H_3 & H_2 + \varepsilon^9 H_6 & 
\varepsilon H_5 + \varepsilon^4 H_1 \\
\varepsilon H_6 + \varepsilon^4 H_2  & 
\varepsilon H_5 + \varepsilon^4 H_1 & 
H_3 + \varepsilon^9 H_4 
\end{array}
\right),
\end{equation}
when all of three sets of D-brane intersect at the same point.

Including many Higgs fields may lead to realistic 
Yukawa matrices for the quark and lepton sectors, 
when we assume proper VEVs for these Higgs fields.
However, that raises another question: how can we 
realize such proper ratios of Higgs VEVs.

\subsection{Asymmetric flavor structure}

Here we study the asymmetric flavor structure on $T^2$, 
that is, the numbers of left and right-handed quarks 
are different each other, e.g. on the first $T^2$, 
i.e. $|I^{(1)}_{CL}| \neq |I^{(1)}_{RC}|$.
The total number of left and right-handed quarks must be 
the same.
Thus, intersecting points, e.g. on the second $T^2$, 
must satisfy $|I^{(1)}_{CL} I^{(2)}_{CL}| = 
|I^{(1)}_{RC} I^{(2)}_{RC}|$, and 
the total flavor number is equal to 
$N_f=|I^{(1)}_{CL}I^{(2)}_{CL}|=|I^{(1)}_{RC} I^{(2)}_{RC}|$.

First, let us investigate which types of D-brane configurations 
can appear.
Again, we consider three sets of D-branes, $D_C$, $D_L$ and $D_R$ branes 
with the following winding numbers on the first $T^2$,
\begin{eqnarray}
D_C &:& {\bf w}^{(1)}_C = (n^{(1)}_C,m^{(1)}_C),\nonumber  \\
D_L &:& {\bf w}^{(1)}_L = (n^{(1)}_L,m^{(1)}_L), \\
D_R &:& {\bf w}^{(1)}_R = (n^{(1)}_R,m^{(1)}_R).  \nonumber
\end{eqnarray}
Suppose that 
\begin{equation}
I_{CL}^{(1)}=\ell C, \qquad I_{RC}^{(1)}=r C,
\end{equation}
where $\ell, r,C \in {\bf Z}$, and $g.c.d.(\ell,r)=1$.
Through a simple algebraic calculation, we can show 
\begin{equation}
\ell {\bf w}_R \pm r {\bf w}_L = j {\bf w}_C,
\end{equation}
where $j $ is integer.
Then, we can calculate 
\begin{equation}
|I_{LR}^{(1)}|=j C.
\end{equation}
Here the integer $j$ must satisfy 
\begin{equation}
g.c.d.(j, \ell) =g.c.d.(j,r)=1 .
\end{equation}
For example, if $g.c.d.(j, \ell) = C' \neq 1$, 
the above discussion could be applied to show that 
$r= C' r'$ with integer $r'$.
That is inconsistent with the above condition 
$g.c.d.(\ell , r)=1$.
Thus, the generic D-brane configuration should satisfy   
\begin{equation}
g.c.d.(I_{CL}^{(1)},I_{RC}^{(1)})=g.c.d.(I_{CL}^{(1)},I_{LR}^{(1)})=
g.c.d.(I_{RC}^{(1)},I_{LR}^{(1)}).
\label{generic-conf}
\end{equation}

We have understood the generic D-brane configuration.
Now, let us consider the selection rule for such generic case.
In section 3, we have shown that all of couplings are allowed 
in the D-brane configuration satisfying eq.(\ref{co-prime}).
Here we extend that into the generic case.
We take $C$ as the greatest common divisor of any two intersecting 
numbers.
Suppose that 
$g.c.d.(n_1^{(1)},n_2^{(1)})=g.c.d.(n_1^{(1)},n_3^{(1)})=
g.c.d.(n_2^{(1)},n_3^{(1)})=1.$
Then, the sublattices are spanned by 
\footnote{See Appendix A.1.}
\begin{eqnarray}
\Lambda_{CL}^{(1)} &:& (1,J_{CL}^{(1)}C) {\rm~~~and~~~} (0,k'_{CL}C), 
\nonumber \\
\Lambda_{RC}^{(1)} &:& (1,J_{RC}^{(1)}C) {\rm~~~and~~~} (0,k'_{RC}C) ,\\
\Lambda_{LR}^{(1)} &:& (1,J_{LR}^{(1)}C) {\rm~~~and~~~} (0,k'_{LR}C) ,
\nonumber
\end{eqnarray}
where 
\begin{equation}
g.c.d.(k'_{CL},k'_{RC})=g.c.d.(k'_{CL},k'_{LR})=
g.c.d.(k'_{RC},k'_{LR})=1 ,
\end{equation}
and $J_{CL}^{(1)}$, $J_{RC}^{(1)}$ and $J_{LR}^{(1)}$ 
are certain integeres, but irrelevant to our discussions.
In this case, the combination of the sublattices, 
$\Lambda_{CL}, \Lambda_{RC}$ and $\Lambda_{LR}$ corresponds to 
the sublattice, which is spanned by 
$(1,0)$ and $(0,C)$.
Therefore, the coupling selection rule 
is determined by the discrete $Z_C$ symmetry.
Let us write this generic result on the coupling selection rule 
in simple words.
When we label intersecting points by $j_{CL}$, $j_{RC}$ and 
$j_{LR}$, i.e., 
\begin{eqnarray}
j_{CL} &=& 0,1,\cdots, I_{CL}^{(1)} -1, \nonumber \\
j_{RC} &=& 0,1,\cdots, I_{RC}^{(1)} -1, \\
j_{LR} &=& 0,1,\cdots, I_{LR}^{(1)} -1, \nonumber
\end{eqnarray}
the coupling selection rule is obtained as 
\begin{equation}
j_{CL} + j_{RC} +  j_{LR} = 0 \quad \pmod{C},
\label{ZC-symmetry}
\end{equation}
where 
\begin{equation}
C=g.c.d.(I_{CL}^{(1)},I_{RC}^{(1)})=g.c.d.(I_{CL}^{(1)},I_{LR}^{(1)})=
g.c.d.(I_{RC}^{(1)},I_{LR}^{(1)}) ,
\label{ZC-rule}
\end{equation}
for $g.c.d.(n_1^{(1)},n_2^{(1)})=g.c.d.(n_1^{(1)},n_3^{(1)})=
g.c.d.(n_2^{(1)},n_3^{(1)})=1$.
This selection rule is mentioned  also in \cite{Cremades:2003qj} 
as Ansatz.

In generic case, the sublattices are spanned by \footnote{
See Appendix A.2. }
\begin{eqnarray}
\Lambda_{CL}^{(1)} &:& (k_{CL}C_1,J^{(1)}_{CL}C_2) 
{\rm~~~and~~~} (0,k'_{CL}C_2), 
\nonumber \\
\Lambda_{RC}^{(1)} &:& (k_{RC}C_1,J^{(1)}_{RC}C_2) 
{\rm~~~and~~~} (0,k'_{RC}C_2) ,\\
\Lambda_{LR}^{(1)} &:& (k_{LR}C_1,J^{(1)}_{LR}C_2) 
{\rm~~~and~~~} (0,k'_{LR}C_2) ,
\nonumber
\end{eqnarray}
where $C_1C_2=C$ and 
\begin{eqnarray}
& & g.c.d.(k_{CL},k_{RC})=g.c.d.(k_{CL},k_{LR})=
g.c.d.(k_{RC},k_{LR})=1, \nonumber \\
& & g.c.d.(k'_{CL},k'_{RC})=g.c.d.(k'_{CL},k'_{LR})=
g.c.d.(k'_{RC},k'_{LR})=1 .
\end{eqnarray}
Then, the combination of the sublattices, 
$\Lambda_{CL}, \Lambda_{RC}$ and $\Lambda_{LR}$ corresponds to 
the sublattice, which is spanned by 
$(C_1,0)$ and $(0,C_2)$.
Hence, the coupling selection rule is determined by 
the $Z_{C_1} \times Z_{C_2}$ symmetry.
This selection rule is also written as a simple extension of 
eq.(\ref{ZC-symmetry}).

We have understood the generic D-brane configuration and its 
coupling selection rule. From phenomenological viewpoint, 
we are interested in D-brane configurations leading to small 
numbers of flavors, in particular the flavor number equal to three.
The D-brane configuration leading to the total flavor number 
$N_f =3$ is realized by the intersecting numbers, 
\begin{equation}
(I^{(1)}_{CL},I^{(1)}_{RC})(I^{(2)}_{CL},I^{(2)}_{RC})
=(3,1)(1,3).
\end{equation}
Suppose that $I^{(1)}_{LR}I^{(2)}_{LR} = 1$, that is, 
the number of Higgs fields is equal to one.
In this case, we have the factorizable form of the Yukawa 
matrix,
\begin{equation}
Y_{ij} = a_ib_j ,
\end{equation}
as already known in the literature.
That is the rank-one matrix, and 
that makes only one of flavor massive, and the others 
remain massless still with obviously vanishing mixing angles.

The minimal case, which can lead to non-vanishing mixing with 
only one Higgs field, 
may be the D-brane configuration with the intersecting numbers,  
\begin{equation}
(I^{(1)}_{CL},I^{(1)}_{RC})(I^{(2)}_{CL},I^{(2)}_{RC})
=(3,2)(2,3).
\end{equation}
The total flavor number $N_f$ is equal to $N_f = 6$, and 
left and right-handed quarks are denoted by 
$Q^{ij}_L$ and $Q^{k \ell}_R$, respectively, 
where the indices $(i,j)$ label intersecting points on 
the first and second tori for $i=0,1,2$ and $j=0,1$.
The indices $(k \ell)$ have the same meaning. 
In this case with one Higgs field, the total Yukawa matrix 
is obtained as a direct product of the parts from the 
first and second tori, i.e.,
\begin{equation}
Y_{(i,j)(k,\ell)} = a_{i k}b_{j \ell}.
\end{equation}
The generic form of $a_{ik}$ and $(b_{j \ell})^T$ is 
obtained by eqs.(\ref{asym-1}), (\ref{asym-2})  
with different suppression factors 
$\varepsilon$ and $\varepsilon'$.
Thus, this Yukawa matrix $Y_{(i,j)(k,\ell)}$ has 
the non-trivial form with the rank-four, that is, 
the mass ratios among massive modes and diagonalizing 
matrix elements are determined by geometrical aspects.
However, when we apply this form to 
up and down  sectors of quarks, it can not realize 
experimental values of masses and mixing angles.
The clear problem is that the flavor number is larger and 
two flavors remain still massless.
Also, another phenomenological aspect is that off-diagonal 
entries are suppressed compared with diagonal entries.
However, experimental values of mass ratios and mixing 
angles in the quark sector as well as the lepton sector satisfy 
$m_i/m_j \leq V_{ij}$ for $i <j$, and that implies 
off-diagonal entries must not be suppressed so much 
in one of up and down sectors.

The generic aspects of the asymmetric flavor structure with the 
intersecting numbers $(I^{(1)}_{CL},I^{(1)}_{RC})(I^{(2)}_{CL},I^{(2)}_{RC})$, 
which satisfy $I^{(1)}_{CL} = I^{(2)}_{RC}$ and 
$I^{(1)}_{RC} = I^{(2)}_{CL}$, is as follows.
Here we consider the case that $g.c.d.(I^{(1)}_{CL},I^{(1)}_{RC})=1$.
The extension to the case with 
$g.c.d.(I^{(1)}_{CL},I^{(1)}_{RC}) \neq 1$ is simple.
We denote left and right-handed quarks by 
$Q^{ij}_L$ and $Q^{k \ell}_R$, respectively, for 
$i, \ell=0,1,\cdots, I^{(1)}_{CL}-1$ and 
$j, k=0,1,\cdots, I^{(1)}_{RC}-1$.
We suppose the minimal number of Higgs field.
In this case, the full Yukawa matrix is obtained as a 
direct product,
\begin{equation}
Y_{(i,j)(k,\ell)} = a_{ik}b_{j\ell} .
\end{equation}
When all of three sets of D-branes intersect at the same 
point, 
the factor matrix corresponding to the first $T^2$, $a_{ik}$, 
is derived through a simple 
calculation,
\begin{equation}
a_{ik} = \varepsilon^{m^2},
\end{equation}
where $m$ is the integer satisfy 
\begin{eqnarray}
i &=& m \quad \pmod{I^{(1)}_{CL}}, \nonumber  \\
k &=& m \quad  \pmod{I^{(1)}_{RC}} ,
\end{eqnarray}
with the minimum $|m|$, 
where $m$ includes negative integers.
This result may be obvious when we draw a 
figure like Fig.~3.
The matrix $b_{j \ell}$ is also obtained in a similar way.
When all of three sets of $D$-brane do not intersect 
at the same point, the factor matrix $a_{ik}$ is obtained 
in the same way except by replacing $m \rightarrow m-d$, 
where $d$ is a continuous parameter $(-1 \leq d \leq 1)$.

The total flavor number $N_f$ is obtained as 
$N_f = I^{(1)}_{CL} I^{(2)}_{CL}$.
The rank of the full Yukawa matrix is equal to 
$I^{(1)}_{CL} I^{(1)}_{CL}$ when $I^{(1)}_{CL} < I^{(2)}_{CL}$.
Thus, the $I^{(1)}_{CL}(I^{(1)}_{RC} - I^{(1)}_{CL})$ modes appear 
massless.
Off-diagonal entries are suppressed compared with 
diagonal entries, and not enough to realize experimental 
values of mixing angles.

Now, let us consider the case with more than one Higgs fields.
We consider the model with the intersecting numbers, 
$(I^{(1)}_{CL},I^{(1)}_{RC})(I^{(2)}_{CL},I^{(2)}_{RC})
=(3,1)(1,3)$.
Even if the intersecting numbers for the Higgs fields are 
$(I^{(1)}_{LR},I^{(2)}_{LR})=(2,1)$ or $(1,2)$, the rank 
of mass matrix is still equal to one.
For example, in the former case, the Yukawa couplings 
for each of Higgs fields $H_a$ $(a=1,2)$ are a factorizable form
like $Y_{ija}=a_{ia} b_j$, and the mass matrix $m_{ij}$ is still a  
factorizable form, 
\begin{equation}
m_{ij} = (a_{i1}\langle H_1 \rangle  + a_{i2}\langle H_2 \rangle )
b_j,
\end{equation} 
that is, the rank-one matrix.
The minimal case increasing the rank of mass matrix is 
$(I^{(1)}_{LR},I^{(2)}_{LR})=(2,2)$, that is, 
the totally four Higgs fields.
Then, the mass matrix becomes rank-two.

To make all of three flavors massive, 
we have to introduce more Higgs fields.
However, recall the rule for D-brane configuration 
discussed in subsection 4.1 when $I^{(1)}_{LR} = 3$ 
or $I^{(2)}_{LR} = 3$.
If we have e.g. $I^{(1)}_{LR} = 3$ in addition to 
$I^{(1)}_{CL} = 3$, such D-brane configuration 
requires at least $I^{(1)}_{RC}=3$, and 
the minimum case with $I^{(1)}_{RC}=3$ corresponds 
to the symmetric flavor structure, which have been 
discussed in the previous subsection.

\subsection{Comments on reducing flavor and 
Higgs numbers}

We have studied the flavor structures, which can be derived from 
intersecting D-brane configurations, including 
the symmetric flavor structures and asymmetric ones.
For the symmetric flavor structure with $N_f=3$, 
corresponding D-brane configurations require 
the number of Higgs fields to be equal to $N_f k$ with integer $k$.
For the asymmetric flavor case, the D-brane configurations 
with $(I^{(1)}_{CL}, I^{(1)}_{RC})(I^{(2)}_{CL}, I^{(2)}_{RC})
=(3,1)(1,3)$ lead to the factorizable form of Yukawa matrix, 
$Y_{ij} = a_ib_j$, that is, the rank-one matrix.
That implies that even if we can realize the massless 
spectrum of the minimal supersymmetric standard model, 
we would have difficulty to derive realistic Yukawa matrices from 
stringy 3-point couplings at the tree-level.\footnote{
Loop corrections have been studied in \cite{Abel:2003yh}.
However, in the softly broken SUSY case, it seems difficult to 
obtain realistic results by loop corrections.}

For the symmetric flavor case as well as the asymmetric case, 
introduction of many Higgs fields may lead to 
realistic Yukawa matrices.
However, in such case, proper values of ratios among 
VEVs of many Higgs fields have to be chosen.
Also, introduction of many Higgs fields may lead to a problem, 
because that, in general, causes flavor changing neutral currents. 
One solution is that we introduce extra fields $H'$, 
which have mass terms with Higgs fields larger than 
the weak scale, and that only one mode remains light.
If such light mode is a linear combination of 
original modes with proper coefficients, 
then realistic Yukawa matrices can be realized.

For the asymmetric flavor case, off-diagonal couplings 
are allowed for one Higgs field, e.g. 
$(I^{(1)}_{CL},I^{(1)}_{RC})(I^{(2)}_{CL},I^{(2)}_{RC})
=(3,2)(2,3)$, and the cases with more intersecting points 
also lead to non-trivial mixing angles.
However, in such case the flavor number is larger than three.
We have to reduce the flavor number, that is, we have to 
introduce anti-generations and mass terms between 
generations and anti-generations.

Thus, it is quite important to generate mass terms 
among Higgs fields $H$ and extra fields $H'$, and/or 
generations and anti-generations of quarks.
One of stringy ways to generate mass terms is  
recombination of D-branes.
However, if those D-branes are stabilized to be bending, 
it is not clear how to treat them.
Otherwise, if those are stabilized not to bend, but be straightened, 
the resultant D-brane configurations would be 
classified into the D-brane configurations, which have 
been studied in subsections 4.1 and 4.2.
Anyway, stringy way to generate mass terms is beyond 
our scope.
Thus, here we give comments on this issue from 
the viewpoint of effective field theory.
Within the framework of effective field theory, 
there may be two ways to generate mass terms; 
one is through symmetry breaking and the other is due to 
compositeness through strong dynamics.

Concerned about the former scenario, 
suppose that we have the following type of couplings,
\begin{equation}
y  H H' X, \qquad y Q Q'X.
\label{extra}
\end{equation}
These couplings are originated from the D-brane configuration 
of Fig. 5 for the Higg fields, and a similar configuration for 
the quarks.
These coupling strengths $y$ are calculated within 
the framework of string theory.
Suppose that the fields $X$ develop their VEVs.
Then, the above operators become effective mass terms between 
$H$ and $H'$, and  $Q$ and $Q'$, and 
that can reduce the number of light flavors and light Higgs fields.
Note that the VEVs of the fields $X$ break gauge symmetries, 
under which $X$ have non-trivial representations.
Thus, the gauge group, which is obtained at the string scale, 
must be larger than one of the standard model.


\begin{figure}[htbp]
\epsfxsize=0.6\textwidth
\centerline{\epsfbox{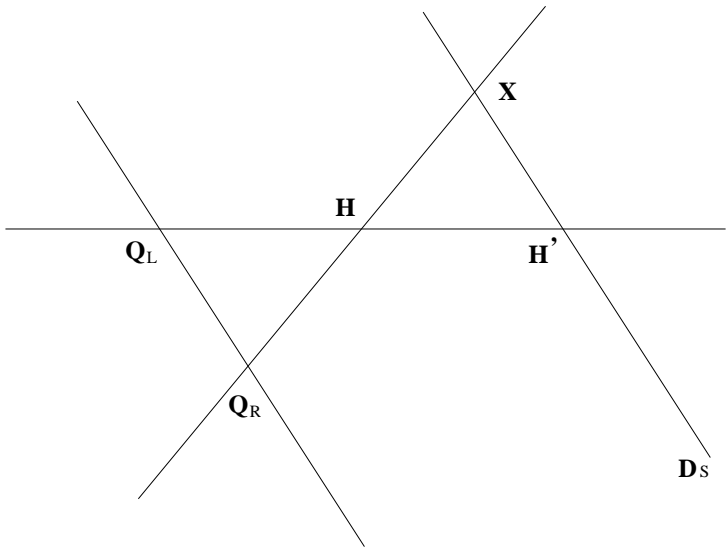}}
\caption{D-brane configuration leading to mass terms between $H$ and 
$H'$.
}
\end{figure}


The latter scenario to generate mass terms is similar 
to the former.
Suppose again that we have the same type of couplings as 
eq.(\ref{extra}).
This time, we assume that the gauge coupling corresponding to 
the gauge group, under which both $X$ and $H'$, and 
$X$ and $Q'$, have non-trivial charges becomes strong.
Such gauge sector corresponds to the $D_s$ brane in Fig.~5.
Then, composite modes appear 
$(H'X)$ and $(Q'X)$.
Obviously, these composite modes have effective mass terms 
with $H$ and $Q$.
Then, the numbers of Higgs fields and flavors can 
be reduced.

In the composite scenario, the light modes can be composite modes, 
when the flavor number from composite modes is  larger than 
one of original modes.
In this case, effective Yukawa couplings can be originated from 
stringy $n$-point couplings for $n= 3,4,5,6$ depending on 
which modes correspond to composite modes.
Actually, such explicit models with composite modes have been 
studied showing an interesting form of Yukawa 
matrices \cite{Kitazawa:2004hz,Kitazawa:2004nf}.
It is quite important to study somehow systematically 
the flavor structure, which can be derived from 
intersecting D-brane configurations, considering 
the above scenarios to reduce the Higgs and flavor numbers 
through the symmetry breaking and strong dynamics.
We have to classify  higher dimensional operators as well as 
3-point couplings.
For such purpose, the discussions in the previous subsections 
would be useful.
However, we leave it for future study.

\section{Conclusion}

We have studied the flavor structure and the coupling selection rule 
within the framework of intersecting D-brane models.
We have formulated the coupling selection rule in terms of 
shift vectors, which are coset representatives  
corresponding to $\Lambda/\Lambda_{ab}$.
With this formulation, we can write the coupling selection rule 
for generic $n$-point couplings in a simple way.

We have found that generic D-brane configurations must 
satisfy the relation (\ref{generic-conf}).
In such generic case, the coupling selection rule is 
determined by the discrete abelian symmetry.
For example, the symmetric flavor structure with $N_f=3$ 
requires at least three Higgs fields, and the coupling selection rule 
is determined by the $Z_3$ symmetry.
We may need more Higgs fields to derive realistic Yukawa matrices.

In the asymmetric flavor structure, the case with $N_f=3$ leads to 
the result that only the third family becomes massive, but the others 
remain massless.
For the asymmetric flavor structure with more flavor, 
we have non-trivial Yukawa 
matrices, although their diagonal entries are quite larger than 
off-diagonal entries and some modes remain still massless.

Our results show that even if we could obtain the minimal matter 
content of the supersymmetric standard model at the string scale, 
we would face with the difficultly to derive realistic Yukawa matrices.
It would be interesting to study alternative scenario that 
we may have several Higgs fields and generations and anti-generations 
of fermions and investigate a way to generate mass terms 
e.g. within the effective field-theoretical way.

We have studied only 3-point couplings at the tree level.
It is important to study higher dimensional operators and 
loop-effects.
In particular, it is interesting to classify D-brane 
configurations with allowed $n$-point couplings for 
$n=3,4,5,6,\cdots$ from the viewpoint of the scenarios discussed 
in subsection 4.3.

In heterotic orbifold models without Wilson lines, 
massless spectra are degenerate on all of fixed points, 
and a large number of quarks and leptons as well as Higgs 
fields appear.
However, Wilson lines can resolve such degeneracy 
\cite{Ibanez:1986tp,Kobayashi:1990mi,Kobayashi:1991rp}, and 
lead to different massless spectra between fixed points.
That is useful to reduce flavor numbers.
We might need such stringy way in intersecting D-brane models.

\section*{Acknowledgment}

N.~K. and T.~K.\/ are supported in part by the Grand-in-Aid for Scientific
Research \#16340078 and \#16028211, respectively.
T.~H.\/, T.~K.\/ and K.~T.\/ are supported in part by 
the Grant-in-Aid for
the 21st Century COE ``The Center for Diversity and
Universality in Physics'' from the Ministry of Education, Culture,
Sports, Science and Technology of Japan.

\appendix

\section{Shift vectors}

In this appendix, we give a simple recipe how to obtain 
shift vectors ${\bf v}_{ab}$ for generic winding numbers, e.g. 
for the first $T^2$.
Here we omit the index (1) for the first $T^2$.
We consider the two sets of D-branes with  the following 
winding numbers,
\begin{eqnarray}
D_a &:& {\bf w}_a^{} = (n_a,m_a^{}), \nonumber \\
D_b &:& {\bf w}_b^{} = (n_b,m_b^{}). 
\end{eqnarray}

\subsection{Case I}

First we consider the case with $g.c.d.(n_a,n_b)=1$.
In this case,  we can show easily 
$g.c.d.(I_{ab},n_a)=g.c.d.(I_{ab},n_b)=1$.
Thus, the set of independent intersecting points 
$\frac{k}{I_{ab}} {\bf w}_a$ $(k=0,1,\cdots, I_{ab} -1)$ 
is equivalent to the sets of 
$\frac{k }{I_{ab}} n_b{\bf w}_a$ $(k=0,1,\cdots, I_{ab} -1)$.
Similarly the  set of independent intersecting points 
$\frac{\ell}{I_{ab}} {\bf w}_b$ $(\ell=0,1,\cdots, I_{ab} -1)$ 
is equivalent to the sets of 
$\frac{\ell }{I_{ab}} n_a{\bf w}_b$ $(k=0,1,\cdots, I_{ab} -1)$.
Obviously, we have 
\begin{equation}
n_a{\bf w}_b - n_b {\bf w}_a = (0,I_{ab}).
\end{equation}
Thus, the sublattice $\Lambda_{ab}$ is spanned by 
$(1,J_{ab})$ and $(0,I_{ab})$, where $J_{ab}$ must be
integer. 
This integer, $J_{ab}$, is irrelevant to description 
of the sets of independent shifts and coset representatives of 
$\Lambda / \Lambda_{ab}$.
When $g.c.d.(m_a,m_b) = C_m \neq 1$, the integer $J_{ab}$ 
is also written as $J_{ab} = J'_{ab} C_m$ with integer $J'_{ab}$.
The shift vectors describing $I_{ab}$ independent 
intersecting points are written as 
\begin{equation}
{\bf v}_{ab} = \frac{k^{(ab)}}{I_{ab}}
(n_a{\bf w}_b - n_b{\bf w}_a )=(0,k^{(ab)}),\qquad 
 (k^{(ab)} = 0,1,\cdots, I_{ab} - 1).
\label{shfit-v-n}
\end{equation}

Similarly, we can obtain the shift vectors ${\bf v}_{ab}$ in 
the case with $g.c.d.(m_a,m_b)=1$.
In this case, the independent shifts are written as 
\begin{equation}
{\bf v}_{ab} = \frac{k^{(ab)}}{I_{ab}}
(m_b{\bf w}_a - m_a{\bf w}_b )=(k^{(ab)},0),\qquad 
 (k^{(ab)} = 0,1,\cdots, I_{ab} - 1).
\label{shfit-v-m}
\end{equation}
When $g.c.d.(n_a,n_b)=g.c.d.(m_a,m_b)=1$, 
both sets of independent shift vectors are equivalent.

\subsection{Case II}

Here we discuss a simple recipe how to obtain shift vectors 
in the case with $g.c.d.(n_a,n_b)=C_n \neq 1$ and 
$g.c.d.(m_a,m_b)=C_m \neq 1$, where $C_n$ and $C_m$ must satisfy 
$g.c.d.(C_n,C_m)=1$.
We write these winding numbers as follows,
\begin{eqnarray}
n_a=n_a'C_n, &\qquad& n_b=n_b'C_n, \\ 
m_a=m_a'C_m, &\qquad& m_b=m_b'C_m, 
\end{eqnarray}
where $n'_{a,b}, C_{n,m} \in {\bf Z}$.
We can do the same discussion as the appendix A.1 except 
replacing 
\begin{equation}
n_{a,b} \rightarrow n'_{a,b}, \qquad 
m_{a,b} \rightarrow m'_{a,b} .
\end{equation}
For example, we obtain 
\begin{equation}
n'_a{\bf w}_b - n'_b {\bf w}_a = (0,\frac{I_{ab}}{C_n})=(0,C_m I'_{ab}), 
\end{equation}
where $I'_{ab} = n'_a m'_b - n'_b m'_a$.
Then, the sublattice $\Lambda_{ab}$ is spanned by 
$(C_n,C_mJ'_{ab})$ and $(0, C_m I'_{ab})$, where 
$J'_{ab}$ must be integer.
This integer, $J'_{ab}$, is irrelevant to coset representatives 
of $\Lambda / \Lambda_{ab}$ like Appendix A.1.
The set of independent shifts are obtained as 
\begin{equation}
{\bf v}_{ab}=(k,k') ,
\end{equation}
where $(k=0,1,\cdots,C_n-1),(k'=0,1,\cdots,\frac{I_{ab}}{C_n}-1)$.
There are other equivalent descriptions of shift vectors.

\section{Symmetric flavor structure with $N_f=2$}

In subsection 4.1, we have shown that if 
$|I_{CL}|=|I_{CR}|=N_f$ on $T^2$, the intersecting number 
$|I_{LR}|$ must satisfy $|I_{LR}| = k N_f$ with $k \in {\bf Z}$, 
that is, the minimum number of $|I_{LR}|$ is equal to 
$N_f$.
Here, again we omit the index for the $i$-th torus like 
subsection 4.1.
However, this statement is not true for $N_f=2$.
Indeed, the case with $N_f = 2$ requires $k \geq 2$.
If the following relation 
\begin{equation}
|I_{CL}|=|I_{CR}|=|I_{LR}|=2 ,
\label{int-no-2}
\end{equation}
is true, we could write 
\begin{equation}
{\bf w}_C = \pm {\bf w}_L \pm {\bf w}_R ,
\label{int-no-2-2}
\end{equation}
where the signs depend on signs of $I_{CL}, I_{CR}$ 
and $I_{LR}$.
However, the relations (\ref{int-no-2}) and 
(\ref{int-no-2-2}) are inconsistent in the D-brane configurations.
The relation (\ref{int-no-2-2}) always leads to 
${\bf w}_C =$ (even,even) except the two cases with 
\begin{eqnarray}
{\bf w}_L &=& ({\rm even, odd}), \qquad 
{\bf w}_R = ({\rm odd, even}), \nonumber \\
{\bf w}_L &=& ({\rm odd, even}), \qquad 
{\bf w}_R = ({\rm even, odd}) .
\end{eqnarray}
However, both of these cases lead to $I_{LR}=$ odd.
Thus,  one can not realize the D-brane configuration 
with the relation (\ref{int-no-2}).

The flavor structure with $N_f =2$ on $T^2$ is not realistic, 
but that can be  useful as a piece of a full flavor structure, 
when we combine it with other flavor structures on $T^2 \times T^2$.
(See section 4.)


\end{document}